\documentclass[12pt]{iopart}
\pdfoutput=1
\usepackage{graphicx}
\usepackage{dcolumn}
\usepackage{bm}
\usepackage{url}
\usepackage{amssymb}
 \expandafter\let\csname equation*\endcsname\relax 
  \expandafter\let\csname endequation*\endcsname\relax 
\usepackage{amsmath}
\usepackage[colorlinks=true, a4paper=true, pdfstartview=FitV,
linkcolor=blue, citecolor=blue, urlcolor=blue]{hyperref}

\include{epsf} 

\newcommand{\N}{{\rm I\! N}}

\begin{document}

\title[Reliability measures of second order SMC applied to wind energy production]{Reliability measures of second order semi-Markov chain applied to wind energy production}
\author{Guglielmo D'Amico}  
\address{Dipartimento di Farmacia, 
Universit\`a `G. D'Annunzio' di Chieti-Pescara,  66013 Chieti, Italy}
\author{Filippo Petroni}
\address{Dipartimento di Scienze Economiche ed Aziendali,
Universit\`a degli studi di Cagliari, 09123 Cagliari, Italy}
\author{Flavio Prattico}
\address{Dipartimento di Ingegneria Industriale e dell'Informazione e di Economia, Universit\`a degli studi dell'Aquila, 67100 L'Aquila, Italy}
\bigskip

\date{\today}

\begin{abstract}
In this paper we consider the problem of wind energy production by using a second order semi-Markov chain in state and duration as a model of wind speed. The model used in this paper is based on our previous work where we have showed the ability of second order semi-Markov process in reproducing statistical features of wind speed. Here we briefly present the mathematical model and describe the data and technical characteristics of a commercial wind turbine (Aircon HAWT-10kW). We show how, by using our model, it is possible to compute some of the main dependability measures such as reliability, availability and maintainability functions. We compare, by means of Monte Carlo simulations, the results of the model with real energy production obtained from data available in the Lastem station (Italy) and sampled every 10 minutes. The computation of the dependability measures is a crucial point in the planning and development of a wind farm. Through our model, we show how the values of this quantity can be obtained both analytically and computationally.
\end{abstract}


\maketitle


\section{Introduction}

Wind is one of the most important renewable energy sources. Wind energy is produced by converting the kinetic energy of wind into electrical energy by means of a generator. For this reason, it is important to dispose of an efficient stochastic model for wind speed changes.\\
\indent First and second order Markov chain models have been extensively used in wind speed modelling and synthetic time series generation, see e.g. Shamshad et al. \cite{sha05},  Nfaoui et al. \cite{nfa04}, Youcef et al. \cite{you03}, Shain et al. \cite{sahin} and Torre et al. \cite{torre}.
\\ \indent The Markovian assumption has, especially in the modeling of wind speed, several flaws. In discrete time Markov chain models, waiting times in a state before making a transition to another state are geometrically distributed. Therefore Markov chains impose artificial assumptions on the structure of the data that, very often, are inappropriate. This leads to a great simplification of the model which is unable to reproduce correctly the statistical properties of the real wind speed process.\\
\indent Semi-Markov chains do not have this constraint, because the waiting time distribution functions in the states can be of any type and this allow the data to speak for themselves without any restriction. For this reason, semi-Markov chains have been extensively applied to different fields \cite{dami09,dami09a,dami09b,jans07,limn01,dami11,dami12a,dami12b}.\\
\indent D'Amico et al. \cite{dapefla1} was the first paper where semi-Markov chains were applied in the modelling of wind speed. In that paper were proposed first and second order semi-Markov models with the aim of generate synthetic wind speed data. It was shown that the semi-Markov models performs better than the Markov chain model in reproducing the statistical properties of wind speed data. In particular, the model recognized as being more suitable is the second order semi-Markov model in state and duration.\\
\indent In this paper we show how to compute dependability measures as availability, reliability and maintainability functions for the second order semi-Markov chain in state and duration. These indicators give important information on the feasibility of the investment in a wind farm by giving the possibility to quantify the uncertainty in the wind energy production.\\
\indent Another important aspect is related to the location of the wind farm. In fact, today, many wind farms are built offshore for different reasons: the wind speed is more powerful and constant due to the absence of obstacles and  visual, environmental and acoustic impact is cut down. The maintenance cost, instead, is higher than the onshore wind farm. A good stochastic model can help the planning of preventive maintenance suggesting when is suitable to execute the maintenance operation. This is possible by analyzing what happens after a particular transition between two different states that is lasted for a certain time period.\\
\indent The results presented here are new and generalize some of the results obtained for semi-Markov chain of order one (see Barbu and Limnios \cite{barb04} and Blasi et al. \cite{Bla04}). The model generalizes also Markov chains and renewal models.\\
\indent We apply our model to a real case of energy production. For this reason, we choose a commercial wind turbine, a 10 kW Aricon HAWT assumed to be installed at the station of L.S.I -Lastem which is situated in Italy.\\
\indent The paper is organized as follows. Section 2 presents some definitions and notation on the second order semi-Markov chain in state and duration. Section 3 describes the database and the technical characteristics of the commercial wind turbine. Section 4 shows the way in which it is possible to compute the dependability measures via kernel transformations and the value computed on the real data and on the synthetic data are compared. In the last section some concluding remarks and possible extensions are presented.

\section{The second order semi-Markov chains in state and duration}

Higher order semi-Markov processes were introduced by Limnios and Oprisan \cite{limn03}. The dependence on the past was given only through past states. These models were recently generalized by D'Amico et al. \cite{dapefla1} that proposed second order semi-Markov chain in state and duration.\\
\indent Let us consider a finite set of states $E=\{1,2,...,S\}$ in which the system can be into and a complete probability space $(\Omega, \emph{F}, \mathbb{P})$ on which we define the following random variables:
\begin{equation}
\label{uno}
J_{n}:\Omega\rightarrow E, \,\,\, T_{n}:\Omega\rightarrow \N .
\end{equation}
\indent They denote the state occupied at the $n$-th transition and the time of the $n$-th transition, respectively. To be more concrete, by $J_{n}$ we denote the wind speed at the $n$-th transition and by $T_{n}$ the time of the $n$-th transition of the wind speed process.\\
\indent We assume that
\begin{equation}
\label{ventuno}
\begin{aligned}
& \mathbb{P}[J_{n+1}=j,T_{n+1}-T_{n}= t |\sigma(J_{s},T_{s}), J_{n}=k, J_{n-1}=i, T_{n}-T_{n-1}=x, 0\leq s \leq n]\\
& \quad =\mathbb{P}[J_{n+1}=j,T_{n+1}-T_{n}= t |J_{n}=k, J_{n-1}=i,T_{n}-T_{n-1}=x ]:=\,\,_{x}q_{i.k,j}(t).
\end{aligned}
\end{equation}
\indent Relation $(\ref{ventuno})$ asserts that, the knowledge of the values $J_{n}, J_{n-1},T_{n}-T_{n-1}$ suffices to give the conditional distribution of the couple $J_{n+1}, T_{n+1}-T_{n}$ whatever the values of the past variables might be. \\
\indent The conditional probabilities $ _{x}q_{i.k,j}(t) $ are stored in a matrix of functions $\mathbf{q}=(_{x}q_{i.k,j}(t))$ called the second order kernel (in state and duration). The element $_{x}q_{i.k,j}(t)$ represents the probability that next wind speed will be $j$ at time $t$ given that the current wind speed is $k$, the previous wind speed state was $i$ and the duration in wind speed $i$ before of reaching wind speed $k$ was equal to $x$ units of time.\\  
\indent We can define the cumulated second order kernel probabilities:
\begin{equation}
\label{ventidue}
\begin{aligned}
& _{x}Q_{i.k,j}(t):=\mathbb{P}[J_{n+1}=j, T_{n+1}-T_{n}\leq t|J_{n}=k,J_{n-1}=i,T_{n}-T_{n-1}=x]\\
& =\sum_{s=1}^{t} {_{x}q_{i.k,j}(s)}.
\end{aligned}
\end{equation}
\indent The process $\{J_{n}\}$ is a second order Markov chain with state space $E$ and transition probability matrix $_{x}\mathbf{P}=\,_{x}\mathbf{Q}(\infty)$. We shall refer to it as the embedded Markov chain.\\
\indent The conditional cumulative distribution functions of the waiting time in each state, given the state previously occupied and the duration of occupancy are defined as
\begin{equation}
\label{ventiquattro}
\begin{aligned}
& _{x}G_{i.k,j}(t)=\mathbb{P}[T_{n+1}-T_{n}\leq t|J_{n}=k,J_{n-1}=i, J_{n+1}=j,T_{n}-T_{n-1}=x]\\
& =\frac{1}{_{x}p_{i.k,j}}\sum_{s=1}^{t}{_{x}q_{i.k,j}(s)}\cdot 1_{\{_{x}p_{i.k,j}\neq 0\}}+1_{\{_{x}p_{i.k,j}=0\}}
\end{aligned}
\end{equation}
\indent Denote by $N(t)=\sup\{n:T_{n}\leq t\}$ $\forall t\in \N$ the number of transitions up to time $t$. The second order semi-Markov chain in state and duration can be defined as
 
 \begin{equation}
 \label{a1}
Z(t)=(Z^{1}(t),Z^{2}(t))=(J_{N(t)-1},J_{N(t)}))
 \end{equation}

\indent If we define, $\forall i,k,j\in E$, and $t\in \N$, the semi-Markov transition probabilities by:
\begin{equation}
\label{venticinque}
_{x}\phi_{i.k,h.j}(t):=\mathbb{P}[J_{N(t)}=j,J_{N(t)-1}=h|J_{0}=k,J_{-1}=i,T_{0}=0,T_{0}-T_{-1}=x],
\end{equation}
\noindent it is possible to prove that they verify the following system of equations:
\begin{equation}
\label{ventisei}
_{x}\phi_{i.k,h.j}(t)=1_{\{i=h,k=j\}} \left(1- \sum_{j \in E} \,  _{x}Q_{i,k;j}(t)\right)+\sum_{r\in
E}\sum_{s =1}^{t}\, _{x}q_{i.k,r}(s)\, _{s}\phi_{k.r,h.j}(t-s).
\end{equation}
\indent For this model more general duration dependent transition probabilities than $(\ref{ventisei})$ have been obtained in D'Amico et al. \cite{dapefla1}.

\section{Database and commercial wind turbine}

As in our previous work \cite{dapefla1} we used a free database of wind speed sampled in a weather station situated in Italy.
The station processes the speed every 10 minute in a time interval ranging from 25/10/2006 to 28/06/2011. 
During the 10 minutes are performed 31 sampling which are then averaged in the time interval.
In this work, we use the sampled data that represents the average of the modulus of the wind speed ($m/s$) without considering a specific direction.
This database is composed of about  230000 wind speed measures ranging from 0 to 16 $m/s$.

In order to apply our semi-Markov model, we discretize wind speed into 8 states (see Table \ref{st}) chosen to cover all the wind speed distribution. This choice is done by considering a trade off between accuracy of the description of the wind speed distribution and the number of parameters to be estimated.  An increase in the number of states better describes the process but requires a larger dataset to get reliable estimates and it could also be not necessary for the accuracy needed in forecasting future wind speeds. On the third column of Table \ref{st} we report the number of times the recorded wind speed was in state $i\in\{1,\dots,7\}$. As it is possible to see the number of occupancies of state 7 is small compared to all other states and wind speed exceeds $8m/s$ in very few cases. We stress that the discretization should be chosen according to the database to be used.
 \begin{table}
\begin{center}

\begin{tabular}{|c|*{3}{c|}|}
     \hline
Sate & Wind speed range $m/s$ & Frequency \\ \hline
1 & 0 to 1 & 1135  \\ \hline
2 & 2 & 55407 \\ \hline
3 & 3 & 92659 \\ \hline
4 & 4 & 49347 \\ \hline
5 & 5 & 18031 \\ \hline
6 & 6 & 7397 \\ \hline
7 & 7 & 3698 \\ \hline
8 & $>$7 & 1874 \\ \hline
\end{tabular} 
\caption{Wind speed discretization}
\label{st} 
\end{center}
\end{table}

\begin{figure}
\centering
\includegraphics[height=7cm]{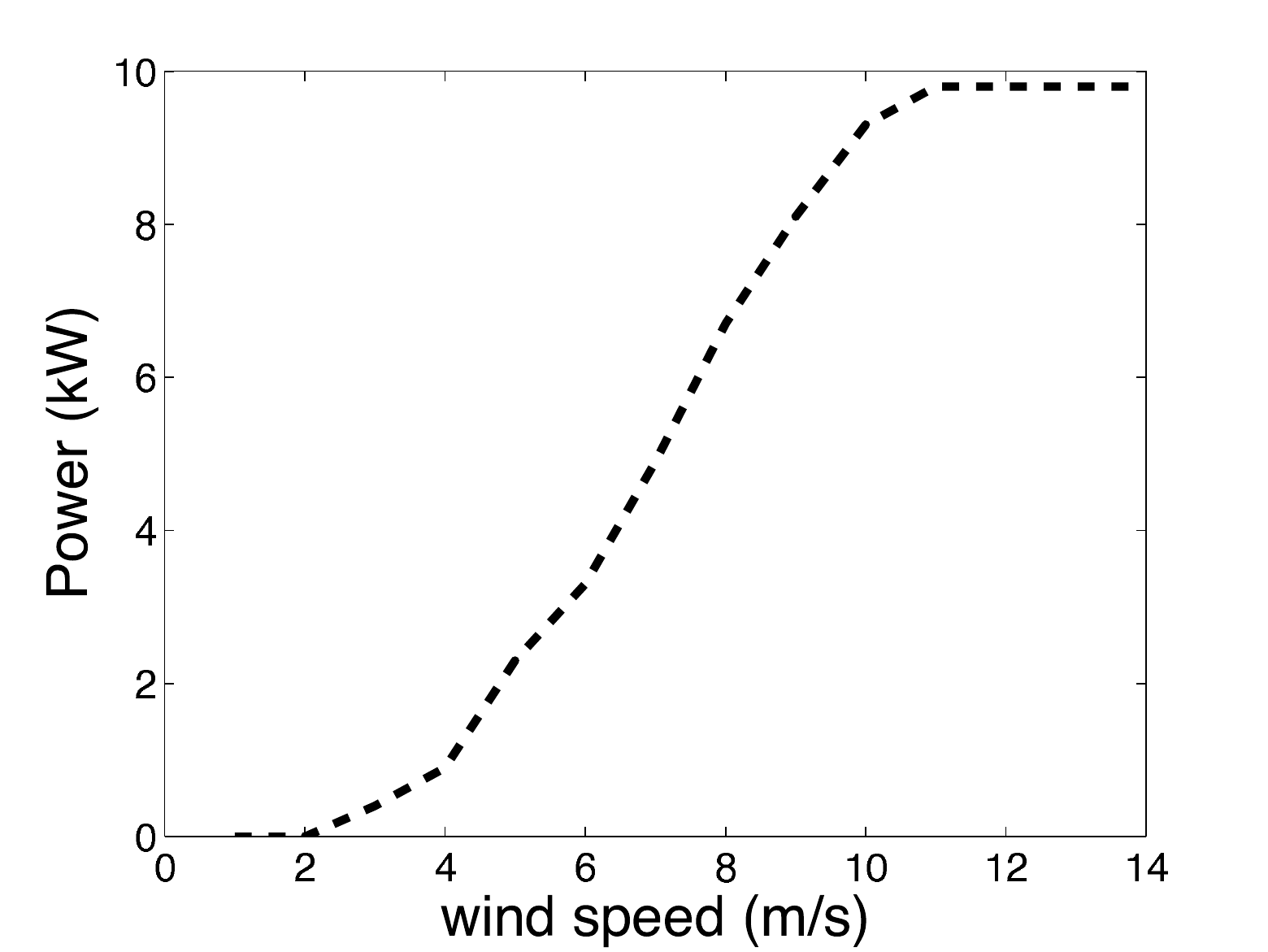}
\caption{Power curve of the 10 kW Aricon HAWT wind turbine.}\label{power}
\end{figure}

We apply our model to a real case of energy production. For this reason we choose a commercial wind turbine, a 10 kW Aricon HAWT with a power curve given in Figure \ref{power}. The power curve of a wind turbine represents how much energy it produces as a function of the wind speed. In this case, see Figure \ref{power}, there is a cut in speed at 2 $m/s$, instead the wind turbine produces energy almost linearly from 3 to 10 $m/s$, then, with increasing wind speed the energy production remains constant until the cut out speed, in which the wind turbine is stopped for structural reason. Then the power curve acts as a filter for the wind speed. In the database used for our analysis the wind speed does never exceed 16 $m/s$ and it is seldom over 8 $m/s$, this is why the discretization is performed according to Table \ref{st} and the wind never reached the cut out speed.

Through this power curve we can know how much energy is produced as a function of the wind speed at a given time.

\section{Reliability theory for the second order semi-Markov chain in state and duration}

In this section, following the research line in Barbu and Limnios \cite{barb04} and in Blasi, Janssen and Manca \cite{Bla04}, we define and compute reliability measures for the second order semi-Markov chain in state and duration.\\
Let $E$ be partitioned into sets $U$ and $D$, so that:
$$
E = U \cup D,\,\,\,\emptyset  = U \cap D,\,\,\,U \ne \emptyset ,\,\,\,U \ne E.
$$
The subset $U$ contains all `Up' states in which the system is working and subset $D$
 all Down' states in which the system is not working well or has failed. In the wind speed model the Up states are those for which the wind speed is sufficiently high to allow the production of energy or not excessive high such that the turbine should be turned off.\\
\indent In the following we present both the typical indicators used in reliability theory and also their application. In order to verify the validity of our model, we compare the behaviour of these indicators for real and synthetic data. The indicators of the synthetic data are computed averaging over 500 different trajectories generated through Monte Carlo simulations based on the second order semi-Markov model in state and duration. The number of trajectories is chosen to have stable results.\\

\indent The three indicators that we evaluate are:

(i) {\it the point wise availability function A} giving the probability that the system is working on at time $t$
 whatever happens on $(0,t]$.\\
 \indent In our model we denote this function by 
\begin{equation}
\label{avalpos}
_{x}A_{i,k}(t) = \mathbb{P}[ Z^{2}(t) \in U| J_{N(0)}=k, J_{N(0)-1}=i,T_{N(0)}-T_{N(0)-1}=x],
\end{equation}
where $Z^{2}(t)=J_N(t)$, see relation $(\ref{a1})$.
The availability $_{x}A_{i,k}(t)$ gives the probability that at time $t$ the wind turbine produces energy given that at time zero the wind speed entered state $k$ coming from state $i$ with a duration equal to $x$.\\

(ii) {\it the reliability function R}
giving the probability that the system was always working from time 0 to time $t$:
\begin{equation}
\label{relpos}
_{x}R_{i,k}(t) = \mathbb{P}[ Z^{2}(u) \in U:\forall u \in ( 0,t] | J_{N(0)}=k, J_{N(0)-1}=i,T_{N(0)}-T_{N(0)-1}=x]
\end{equation}
The reliability $_{x}R_{i,k}(t)$ gives the probability that the wind turbine will always produce energy from time zero up to time $t$ given that at time zero the wind speed entered state $k$ coming from state $i$ with a duration equal to $x$.\\
(iii) {\it the maintainability function M} giving the probability that the system will leave the set $D$
 within the time $t$ being in $D$ at time 0:
\begin{equation}
\label{mantpos}
_{x}M_{i,k}(t) = 1 - \mathbb{P}[ Z^{2}(u) \in D,\,\,\,\forall u \in (0,t] | J_{N(0)}=k, J_{N(0)-1}=i,T_{N(0)}-T_{N(0)-1}=x].
\end{equation}
The maintainability $_{x}M_{i,k}(t)$ gives the probability that the turbine will produce energy at least once from time zero up to time $t$ given that at time zero the wind speed entered state $k$ coming from state $i$ with a duration equal to $x$.\\
\indent These three probabilities can be computed in the following way if the process is a second-order semi-Markov chain in state and duration of cumulated kernel ${\mathbf{Q}}=(_{x}Q_{i,k;j}(t))$.

The three indicators, computed and showed in the following figures, are plotted by varying the initial state \textit{i}, the current state \textit{k} and the sojourn time \textit{x}. The numeric choice of each parameter is given only for graphical reasons, in order to show the maximum number of curves without overlaps. As numeric indicator to compare the gap between the curves we compute the mean square error (MSE) between the indicator applied to the real data and the 500 simulated trajectories.
The mean square error is defined as follows:
\begin{equation}
\label{mse}
_{x}MSE_{i,k}(t)= \sqrt{\frac{1}{500}\sum_{h=1}^{500}\big(_{x}I_{i,k}^{real}(t)-_{x}I_{i,k}^{h}(t)\big)^{2}}
\end{equation}
where $_{x}I_{i,k}^{real}(t)$ stands for the indicators $_{x}A_{i,k}(t)$, $_{x}R_{i,k}(t)$ or $_{x}M_{i,k}(t)$ estimated form real data and $_{x}I_{i,k}^{h}(t)$ the indicators $_{x}A_{i,k}(t)$, $_{x}R_{i,k}(t)$ or $_{x}M_{i,k}(t)$ estimated from each synthetic trajectory.

(i) {\it the point wise availability function}
$
_{x}A_{i,k}(t):
$\\
to compute these probabilities it is sufficient to use the following formula:
\begin{equation}
\label{avaleq}
_{x}A_{i,k}(t) = \sum_{h \in E}\sum_{j \in U} \; _{x}\phi _{i,k;h.j} (t) .
\end{equation}

\begin{figure}
\centering
\includegraphics[height=8cm]{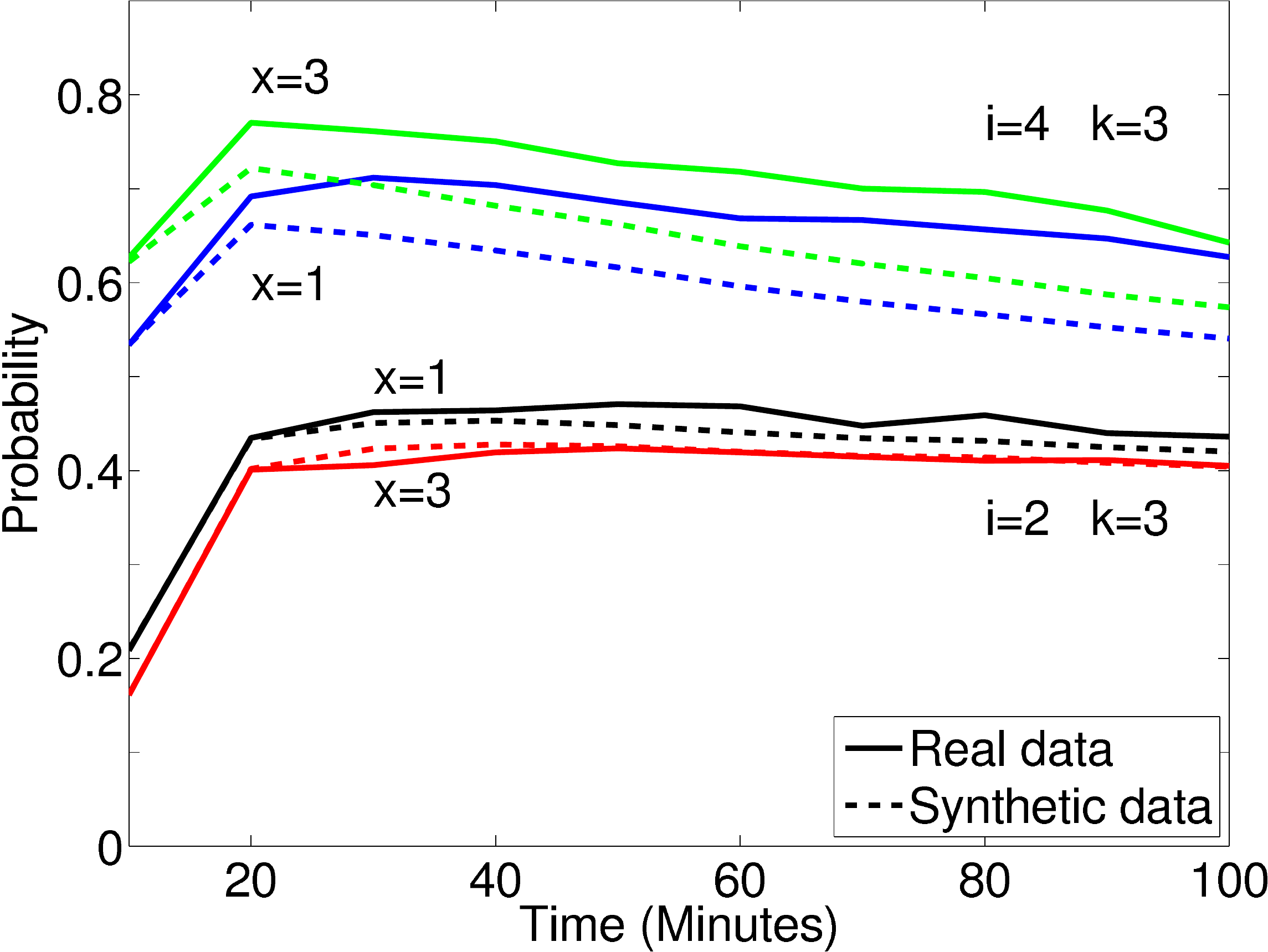}
\caption{Comparison of availability functions for real and simulated data}\label{av}
\end{figure}
\begin{table}
\begin{center}

\begin{tabular}{|c|*{3}{c|}}
     \hline

 & 10 $min$ & 50 $min$  & 100 $min$ \\ \hline
$ _{3}MSE_{4,3} $ & 0.0216 & 0.0679 & 0.0727  \\ \hline
$ _{1}MSE_{4,3} $ & 0.0095 & 0.0700 & 0.0876 \\ \hline
$ _{1}MSE_{2,3}$ & 0.0063 & 0.0180 & 0.0288 \\ \hline
$ _{3}MSE_{2,3} $ & 0.0106 & 0.0135 & 0.0140 \\ \hline

\end{tabular} 
\caption{Mean square error between the curves of the availability applied to real and synthetic data}
\label{eav} 
\end{center}
\end{table}

In Figure \ref{av} the availability functions of the real and synthetic data are compared. The comparison is made by varying the sojourn time and the starting state. Particularly, in Figure \ref{av} the availability function for two different initial states \textit{i} is plotted maintaining constant the current state \textit{k} for two different sojourn times \textit{x}. In Table \ref{eav} we show the MSE for each of the curves of Figure \ref{av}.

(ii) {\it the reliability function}
 $
_{x}R_{i,k}(t):
$

\noindent to compute these probabilities, we will now work with another cumulated kernel ${\hat{\mathbf{Q}}}=(_{x}\hat{Q}_{i,k;j}(t))$
for which all the states of the subset $D$ are changed into absorbing states by considering $ \forall i \in E $ the following transformation:
\begin{equation}
\label{transrelia}
_{x}\hat{p}_{i,k;j}  = {\text{ }}\left\{ {\begin{array}{*{20}c}
   {_{x}p_{i,k;j} \,\,\,\,\,\,{\text{if}}\,\, k\in U,\, \forall j \in E }  \\
   {1 \,\,\,\,\,\,{\text{if}}\,\, k \in D,\, j=k }  \\
   {0\,\,\,\,\,\,\,{\text{if}}\,\, k \in D,\, j \neq k }  \\
 \end{array} } \right.
\end{equation}
$
_{x}R_{i,k}(t)
$
 is given by solving the evolution equation of a second-order semi-Markov chain in state and duration but now with the cumulated kernel $_{x}\hat{Q}_{i,k;j}(t)=_{x}\hat{p}_{i,k;j}\cdot \, _{x}G_{i,k;j}(t)$.\\ 
\indent The related formula will be:
\begin{equation}
\label{releq}
_{x}R_{i,k}(t) = \sum_{h\in U}\sum\limits_{j \in U}   {_{x}\hat{\phi} _{i,k;h.j} (t)}
\end{equation}
where 
$
_{x}\hat{\phi} _{i,k;h.j} (t)$
are the transition probabilities of the process with all the states in 
$
D
$
 that are absorbing, i.e. with cumulated kernel ${\hat{\mathbf{Q}}}$.
 
\begin{figure}
\centering
\includegraphics[height=8cm]{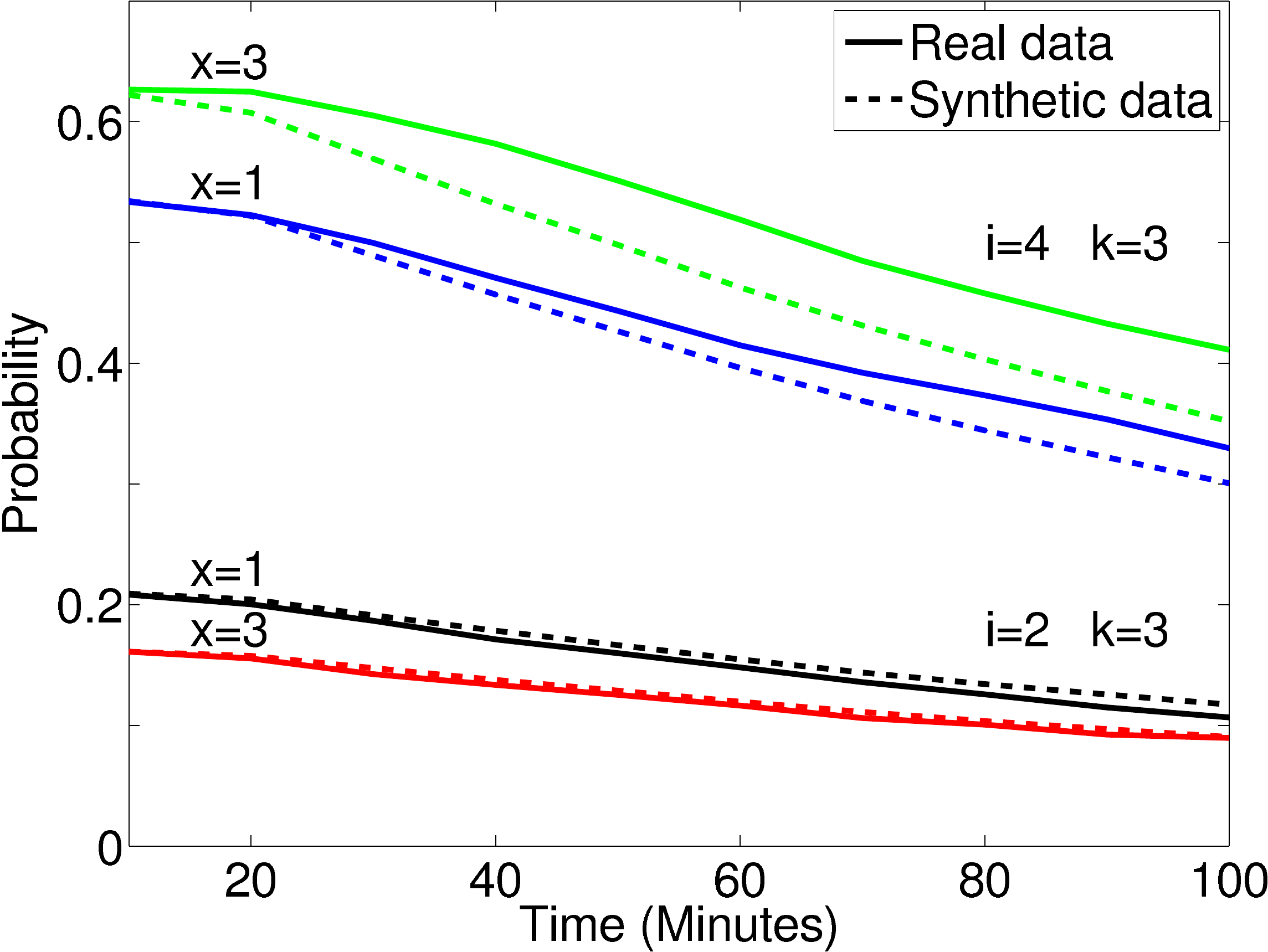}
\caption{Comparison of reliability functions for real and simulated data}\label{re}
\end{figure}

Figure \ref{re} shows the reliability functions for real data compared with the simulated ones. The plotting procedure is the same as for the previous figure: we maintain constant the current state $k$ and we vary the initial state $i$ and the sojourn time $x$. The theoretical trend of the reliability function is confirmed, the probability decreases at the increasing of the time interval. A numerical comparison is given in Table \ref{ere} in which the MSE of the four curves of Figure \ref{re} is computed.

\begin{table}
\begin{center}

\begin{tabular}{|c|*{3}{c|}}
     \hline

 & 10 $min$ & 50 $min$  & 100 $min$ \\ \hline
$ _{1}MSE_{4,3} $ & 0.0095 & 0.0198 & 0.0310 \\ \hline
$ _{3}MSE_{4,3} $ & 0.0216 & 0.0571 & 0.0627 \\ \hline
$ _{1}MSE_{2,3}$ & 0.0063 & 0.0087 & 0.0116 \\ \hline
$ _{3}MSE_{2,3} $ &  0.0016 &0.0097 & 0.0121 \\ \hline

\end{tabular} 
\caption{Mean square error between the curves of the reliability applied to real and synthetic data}
\label{ere} 
\end{center}
\end{table}

(iii) {\it the maintainability function} $_{x}M_{i,k}(t)$:\\
\noindent to compute these probabilities we will now work with another cumulated kernel ${\tilde{\mathbf{Q}}}=(_{x}\tilde{Q}_{i,k;j}(t))$
for which all the states of the subset $U$ are changed into absorbing states by considering the following transformation:
\begin{equation}
\label{transmain}
_{x}\hat{p}_{i,k;j}  = {\text{ }}\left\{ {\begin{array}{*{20}c}
   {_{x}p_{i,k;j} \,\,\,\,\,\,{\text{if}}\,\, k\in D,\, \forall j \in E }  \\
   {1 \,\,\,\,\,\,{\text{if}}\,\, k \in U,\, j=k }  \\
   {0\,\,\,\,\,\,\,{\text{if}}\,\, k \in U,\, j \neq k }  \\
 \end{array} } \right.
\end{equation}
$
_{x}M_{i,k}(t)
$
 is given by solving the evolution equation of a second-order semi-Markov chain in state and duration but now with the cumulated kernel $_{x}\tilde{Q}_{i,k;j}(t)=_{x}\tilde{p}_{i,k;j}\cdot _{x}G_{i,k;j}(t)$.\\ 
 \indent The related formula for the maintainability function will be:
\begin{equation}
\label{manteq}
_{x}M_{i,k}(t) = \sum_{h\in E}\sum\limits_{j \in U} {_{x}\tilde{\phi}_{i,k;h.j} (t)}
\end{equation}
where 
$
_{x}\tilde{\phi}_{i,k;h.j} (t)
$
  is the transition probability of the process with all the states in 
$
U
$
 that are absorbing, i.e. with cumulated kernel ${\tilde{\mathbf{Q}}}$.\\

\begin{figure}
\centering
\includegraphics[height=8cm]{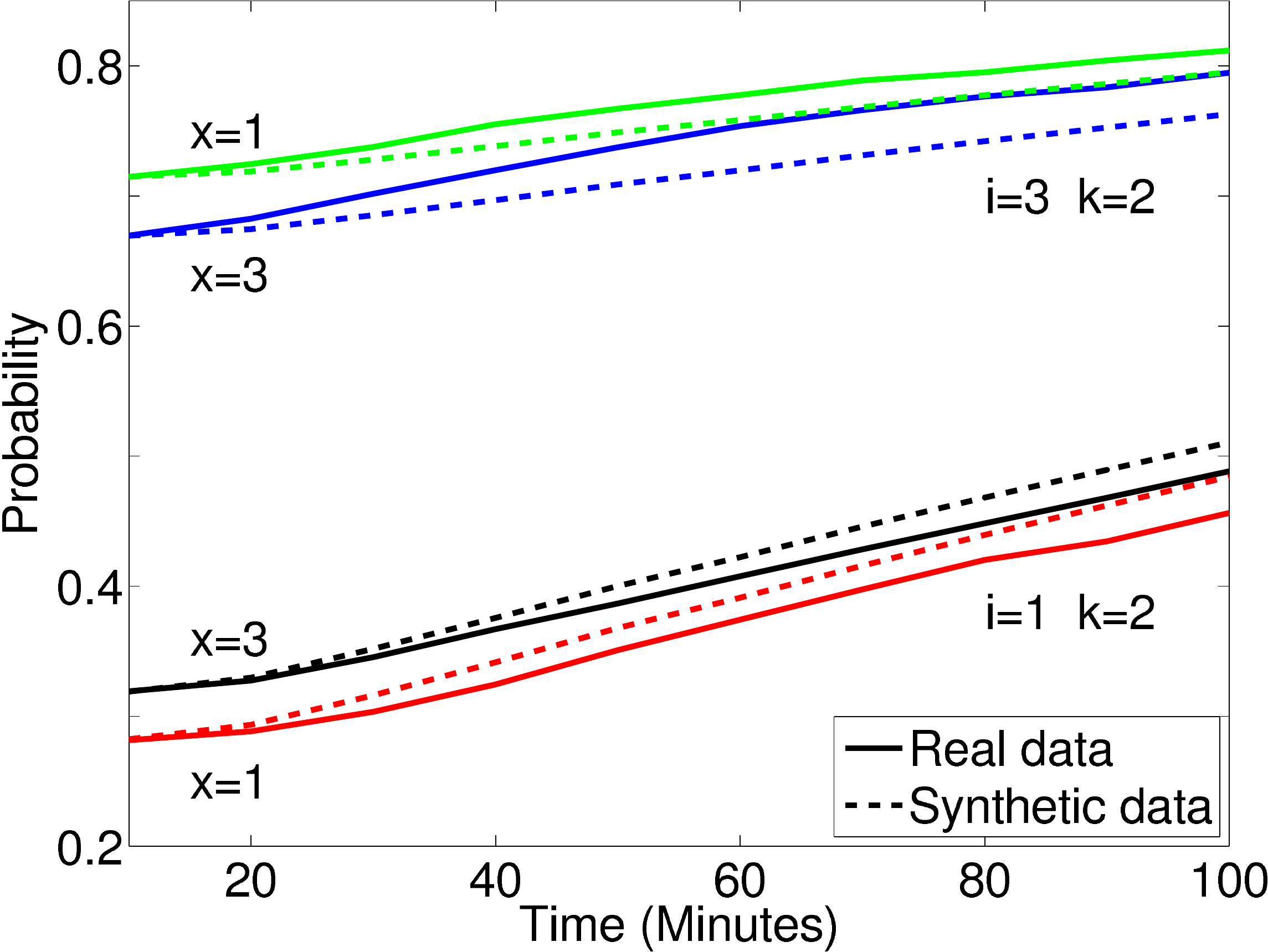}
\caption{Comparison of maintainability functions for real and simulated data}\label{ma}
\end{figure}

The maintainability function is plotted in Figure \ref{ma}. As the previous figures, this one shows the comparison of the maintainability for the real and simulated data varying the initial state and sojourn time and maintaining constant the current state. In Table \ref{ema} the MSE of the maintainability function for the curves of Figure \ref{ma} is computed.

\begin{table}
\begin{center}

\begin{tabular}{|c|*{3}{c|}}
     \hline

 & 10 $min$ & 50 $min$  & 100 $min$ \\ \hline
$ _{1}MSE_{3,2} $ & 0.0058 &  0.0293 & 0.0322\\ \hline
$ _{3}MSE_{3,2} $ & 0.0119 & 0.0199 & 0.0215 \\ \hline
$ _{1}MSE_{1,2}$ & 0.0064 & 0.0153 & 0.0236  \\ \hline
$ _{3}MSE_{1,2} $ & 0.0134 & 0.0220 & 0.0314  \\ \hline

\end{tabular} 
\caption{Mean square error between the curves of the maintainability applied to real and synthetic data}
\label{ema} 
\end{center}
\end{table}

It is possible to note that all the indicators plotted above (availability, reliability and maintainability) depend strongly on the initial and current states and that there is also a great dependence on the sojourn time $x$. In fact, all the probabilities have different values also if only the sojourn time $x$ is changed keeping constant initial states $i$ and final state $k$. For example, from Figure \ref{re} it is possible to see that, in general 
$$
_{1}R_{4,3}(s) > \,_{3}R_{4,3}(s) \, \forall s\in [0,100],
$$
and in particular $_{1}R_{4,3}(40)=0,582$ and $_{3}R_{4,3}(40)=0,473$. This reveals that it is important to dispose of a model that is able to distinguish between these different situations which are determined only from a different duration of permanence in the initial state $i$ before making a transition to the current state $k$. Models based on Markov chains or classical semi-Markov chain are unable to capture this important effect that our second order semi-Markov chain in state and duration reproduces according to the real data.\\  

For each of the indicator we have computed the mean square error between real and synthetic data at different time distance from time zeros. In all cases, the difference between real and simulated data increases with time. The data we are using are 10  minutes sampled, this implies that our model works well for very short time scales and its performance becomes worse for longer time scale. To get better results on a longer time horizon (from more than 1 hour to days) one should use data sampled less frequently but for a longer period. Then, given that the transition probabilities in our model are estimated for high frequency date it is intended for forecasting the probabilities of reliability measures at very short time scale (from 10 minutes to 1 hour).

\section{Conclusion}
In our previous work, we presented new stochastic models for the generation of synthetic wind speed data. In this work, instead, we compute, for the first time, typical indicators in reliability theory for wind speed phenomenon by using a second order semi-Markov model in state and duration. In order to check the validity of the presented model, we have compared the behaviour of these indicators for real data and data simulated by means of Monte Carlo simulations. To do this, we applied our model to a real case of energy production, filtering real and simulated data by the power curve of a commercial wind turbine. 
To compare results from real and simulated date we have computed the mean square error between real and synthetic data for each of the indicator. The results show that the proposed model is able to reproduce the behaviour of real data by exhibiting the dependence of the reliability indicators on past visited states and on the length of the sojourn times. This shows that semi-Markov approach is more suitable than simpler Markov chain models. From our results we can also say that our model applied to 10 minutes sampled data works well if one considers short time scale (from 10 minute to 1 hour) and its performance decreases with time.\\
\indent The indications provided by the model are of importance for assessing the suitability of a location for the wind farm installation as well as for the planning of a preventive maintenance policy.

\end{document}